# Time delay anisotropy in photoelectron emission from the isotropic ground state of helium

Sebastian Heuser[1*], Álvaro Jiménez Galán[2], Claudio Cirelli[1], Mazyar Sabbar[1], Robert Boge[1], Matteo Lucchini[1], Lukas Gallmann[1,3], Igor Ivanov[4,5], Anatoli S. Kheifets[4], J. Marcus Dahlström[6,7,8], Eva Lindroth[6], Luca Argenti[2], Fernando Martín[2,9,10], Ursula Keller[1]

[1]*Physics Department, ETH Zurich, 8093 Zurich, Switzerland*

[2]*Departamento de Química, Módulo 13, Universidad Autónoma de Madrid, 28049 Madrid, Spain*

[3]*Institute of Applied Physics, University of Bern, 3012 Bern, Switzerland*

[4]*Research School of Physics and Engineering, The Australian National University, Canberra ACT 0200, Australia*

[5]*Center for Relativistic Laser Science, Institute for Basic Science (IBS), Gwangju 500-712, Republic of Korea*

[6]*Department of Physics, Stockholm University, AlbaNova University Center, SE-10691 Stockholm, Sweden*

[7]*Max Planck Institute for the Physics of Complex Systems, Noethnitzerstr. 38, 01187 Dresden, Germany*

[8]*Center for Free-Electron Laser Science, Luruper Chaussee 149, 22761 Hamburg, Germany*

[9]*Instituto Madrileño de Estudios Avanzados en Nanociencia (IMDEA-Nanociencia), Cantoblanco, 28049 Madrid, Spain*

[10]Condensed Matter Physics Center (IFIMAC), Universidad Autónoma de Madrid, 28049 Madrid, Spain

*e-mail: sheuser@phys.ethz.ch

**Abstract**

Time delays of electrons emitted from an isotropic initial state and leaving behind an isotropic ion are assumed to be angle-independent. Using an interferometric method involving XUV attosecond pulse trains and an IR probe field in combination with a detection scheme, which allows for full 3D momentum resolution, we show that measured time delays between electrons liberated from the $1s^2$ spherically symmetric ground state of helium depend on the emission direction of the electrons relative to the linear polarization axis of the ionizing XUV light. Such time-delay anisotropy, for which we measure values as large as 60 attoseconds, is caused by the interplay between final quantum states with different symmetry and arises naturally whenever the photoionization process involves the exchange of more than one photon in the field of the parent-ion. With the support of accurate theoretical models, the angular dependence of the time delay is attributed to small phase differences that are induced in the laser-driven continuum transitions to the final states. Since most measurement techniques tracing attosecond electron dynamics involve the exchange of at least two photons, this is a general, significant, and initially unexpected effect that must be taken into account in all such photoionization measurements.

## Introduction

The advent of attosecond science[1,2] (1 as = $10^{-18}$ s) paved the way towards studying and understanding the nature of electron dynamics in atomic, molecular and condensed matter systems on their natural timescale[3-5]. In particular, recent experimental studies in atomic systems[3,6,7] confirmed the ability of attosecond science to unravel ultrafast electron dynamics with high accuracy. A series of groundbreaking investigations[8-14] have established attosecond technology as a new indispensable tool in atomic, molecular, and optical physics.

Extremely small delays in electron emission induced by single photon atomic absorption have been measured with two different techniques such as attosecond energy streaking[15] and RABBITT (reconstruction of attosecond beating by interference of two-photon transitions)[16]. These methods are based on single photoionization, realized in a non-sequential pump-probe scheme where the extreme ultraviolet (XUV) attosecond pump pulse ionizes the target system and an infrared (IR) probe pulse interacts with the liberated electrons. While attosecond streaking employs a single attosecond pulse (SAP), an attosecond pulse train (APT) is used in RABBITT. Neither technique gives access to absolute photoemission time delays. However, relative timing information between electrons originating from different states within the same atom[3,6] or from different atoms[17-19] can be extracted.

An alternative perspective on the photoemission process can be obtained by studying the relative timing of electrons emitted from the same initial state within the same target system but at different emission angles $\theta$, relative to the polarization axis of the XUV-pump (Fig. 1a).

With the attosecond energy streaking technique the emission of electrons is normally only recorded along the linear polarization axis of the IR field, for which the streaking of the photoelectron momentum features a pronounced single-sweep per laser period ($\omega$-modulation). As the ejection-angle is changed from 0° to 90° relative to the linear polarization of the IR field, the streaking of the photoelectron momentum changes to a much weaker $2\omega$-modulation[15]. To avoid any mixing of electrons emitted at different angles, therefore, one must significantly decrease the IR streaking field intensity, which renders the analysis of the experimental streaking traces complex and demanding[20].

With the RABBITT technique the directionality of the momentum transfer is instead of minor importance because the so-called sideband (SB) signals exhibit a $2\omega$-modulation for all ejection angles. Therefore, this method is better suited to explore the angular dependence of observables such as photoemission time delays.

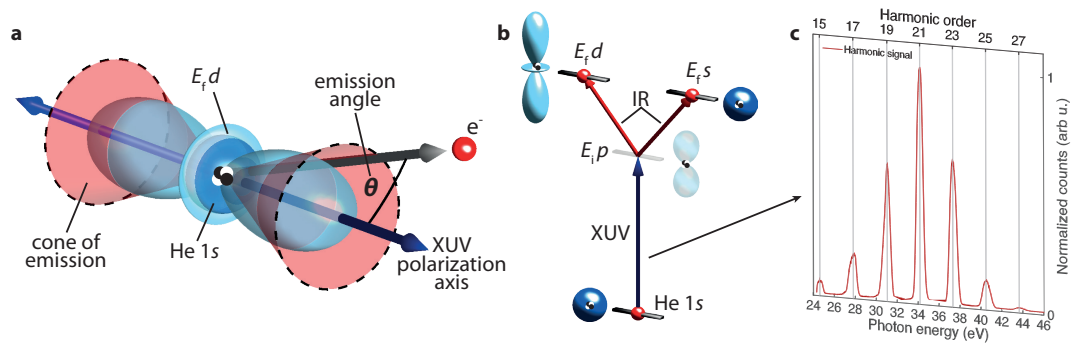

**Figure 1 | Two-photon ionization pathways starting from ground state helium. a,** Schematic defining the emission angle $\theta$ as the electron emission direction relative to the XUV-pump polarization axis and illustrating the different photoelectron partial waves of the corresponding final quantum states, which arise from the exchange of two photons. **b,** Schematic illustrating the different quantum paths, which contribute to the final state of the liberated photoelectrons after the interaction with the XUV and IR fields. **c,** XUV spectrum, which has been used to carry out the experiments.

In atomic photoionization, the absorption of a single photon causes electrons to be excited from their initial state $n_i\, l_i$ into the final state $El$. Here, $n(l)$ is the principal (orbital angular momentum) quantum number and $E$ is the photoelectron energy. Since a photon itself carries a spin angular momentum of one, the allowed transitions $n_i l_i \rightarrow El$ upon the absorption of a single photon come along with a change in angular momentum of $\Delta l = \pm 1$ and therefore two final quantum states with $l \rightarrow l \pm 1$ become accessible. As shown in earlier work[21,22], the interplay between two different angular components may give rise to anisotropic group delays $\tau_W$ (also known as Wigner time delay)[23] of the photoelectron wave packet, which is generated by the absorption of one XUV photon. So far, such an angular dependence was exclusively studied in the context of the ionization from a non-symmetric orbital, and assuming that the transition promoted by the IR did not induce any additional dependence on the emission angle[24].

In the special case of starting from a spherically symmetric orbital $ns$, however, only a single photoionization transition $l \rightarrow l+1$ (i.e., $ns \rightarrow Ep$) is possible. If in addition the remaining ion is left in a spherically symmetric state, the orbital angular momentum of the photoelectron is conserved. In these conditions, the Wigner time delay is rigorously independent on the ejection angle, and so would be the time delay measured with an attosecond interferometric technique, provided that the further exchange of an IR photon did not induce additional angular modulations. Yet, as soon as two photons are involved in the ionization process, two different final states $1s \rightarrow E_i p \rightarrow E_f s / E_f d$ become accessible (Fig. 1b). As a result, in principle, the group delay of the final photoelectron wave packet may still exhibit an angular dependence. This would be the case, for example, if helium (He) was ionized from its spherically symmetric ground state. Indeed, while one expects an isotropic photoemission time delay associated to the XUV absorption, a perturbative analysis (see supplementary material) shows that the intrinsic two-photon nature of the interferometric measurement of the time-delay introduces by itself an inherent, universal anisotropy in the measurement. To which extent such anisotropy affects measurements of photoemission time delays along fixed directions, therefore, is a fundamental question of current attosecond spectroscopy, which has not been addressed before, due in part to the formidable challenges that angle-resolved measurements of photoemission time delays entail.

Here, we present a rigorous experimental and theoretical investigation of the angle-dependent photoemission time delay of electrons removed from the spherically symmetric $^1S^e(1s^2)$ ground state of He to produce the spherically symmetric ion $He^+(1s)$. Full angular resolution is obtained with the recently developed "AttoCOLTRIMS" apparatus[25], which consists of a reaction microscope allowing for full 3D momentum detection[26], combined with an attosecond front-end providing XUV attosecond pulses. Using the RABBITT technique, we measure a significant angular dependence of the photoionization time delay, which can be as large as 60 attoseconds, thus highlighting a new general aspect of electron dynamics triggered by ultrashort pulses in attosecond measurements. We trace back such anisotropy to the electrostatic potential of the parent ion, which influences the transitions induced by the probe pulse. The angle-resolved measurement of photoemission time delays, therefore, is a particularly sensitive tool to probe short-range effects.

**Main text**

In RABBITT spectroscopy, an APT with photon energies in the XUV range (Fig. 1c) is used in combination with an IR probe pulse to trace the electron dynamics by recording the electron kinetic energy as a function of the pump-probe delay $\tau$. In the frequency domain, an APT is formed by odd multiples of the fundamental frequency $\omega_{IR}$ of the driving IR laser pulses employed for high-harmonic generation. Therefore, photoelectrons extracted with an APT from the ground state of an atomic target, with ionization energy $I_p$, are promoted into the continuum at energies $E_{elec} = E_{harm} - I_p$, which mirror the discrete harmonic energies $E_{harm} = (2q+1) \cdot \hbar\omega_{IR}$ of the exciting XUV spectrum, where $q$ is an integer.

The subsequent interaction of the photoelectrons with the weak IR-probe field allows an additional absorption or emission of an IR photon such that also energies corresponding to even multiples of the fundamental frequency become accessible. Thus, SBs in between two consecutive harmonics in the photoelectron spectrum appear. For each SB of order $2q$, there are two indistinguishable excitation pathways: (1) absorption of one photon form harmonic $2q$-1 followed by the *absorption* of an additional IR photon, and (2) absorption of one photon from harmonic $2q$+1 and subsequent *emission* of an IR photon. These two quantum paths interfere, leading to an oscillation of the SB amplitude when changing the delay $\tau$ between the APT and the IR pulse: $SB \propto \cos(2\omega_{IR}\tau - \Delta\phi_{atto} - \Delta\phi_{atomic})$. Here, $\Delta\phi_{atto}$ is the phase difference between consecutive harmonics and corresponds to the group delay of the APT, $\tau_{atto} \approx \Delta\phi_{atto} / 2\omega_{IR}$, while $\Delta\phi_{atomic}$ corresponds to the so-called atomic time delay $\tau_{atomic} \approx \Delta\phi_{atomic} / 2\omega_{IR}$.

Theoretical models[27,28] established that the atomic delay measured along the polarization direction, $\tau_{atomic}$, can be divided into two contributions: the Wigner delay, $\tau_W$, originating from the single-photon XUV ionization[29] and a measurement-induced component, $\tau_{CC}$, which arises due to the additional quantum transition between two electronic states in the continuum induced by the IR-probe pulse in the presence of the Coulomb potential of the ion.

To date the possible dependence of $\tau_{CC}$ on the photoemission angle has not been considered because all measurements on photoionization time delays have used either angle-integrating detection schemes, such as a magnetic bottle spectrometer[6], or directional detectors using e.g. time-of-flight spectrometers[3].

In our RABBITT experiments we have access to all electron emission angles relative to the common XUV/IR polarization axis within one single measurement, thus under the same experimental conditions. Therefore, the contribution from $\tau_{atto}$, which is the same for all the electrons, cancels and we have direct access to the relative atomic delay differences $\Delta\tau_{atomic}$ between electrons emitted at different angles.

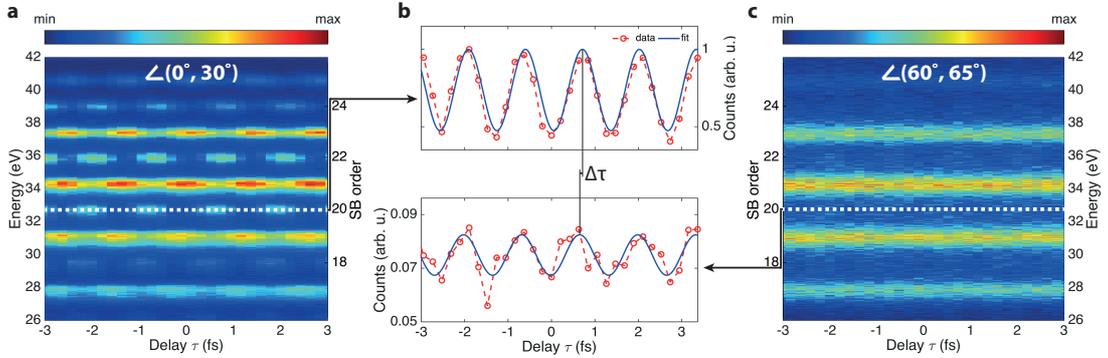

**Figure 2 | Principle of the time delay extraction. a-c**, Examples of measured RABBITT spectrograms and oscillations of sideband (SB) 20 (marked by white dashed lines) for different ranges in emission angle. Note that the energy scale corresponds to the sum of electron kinetic energy and the ionization potential of helium (ionization potential of He: 24.5874 eV). In (**a**) only the electrons detected within a 30° cone of emission (Fig. 1a) are selected. Panel (**c**) comprises electrons emitted within a cone of emission between 60° and 65°. Panel (**b**) shows an example of the intensity oscillations of SB 20 (red data points) obtained by integrating the counts within an energy window of 0.75 eV centered at the peaks of the SB oscillations (white dashed lines) together with their corresponding fits (blue solid lines). The time delay $\Delta\tau$ is clearly visible as a temporal shift between the two different SB oscillations.

To reveal the fundamental angular dependence of $\tau_{CC}$, we performed our investigation with He because $\tau_W$ is in this case rigorously isotropic and He is the only atomic system fully accessible to theory (i.e. apart from atomic hydrogen which is much more challenging for experiments). Therefore this fundamental study can be used as a benchmark.

Figure 2 shows the principle of the angle-resolved RABBITT measurements. Applying an angular filter on the detected electrons, i.e., choosing electrons emitted within the corresponding cone of emission (Fig. 1a), we are able to obtain distinct RABBITT traces representing only electrons out of particular hollow cones (Figs. 2a, 2c). For any angular sector, the SB signal is obtained by integrating the spectrogram in an energy window $\Delta E$=0.75 eV centered at the peak of the SB position. Two curves showing the SB signal are presented in Fig. 2b for electrons emitted between 0° and 30° (top panel) and between 60° and 65° (lower panel).

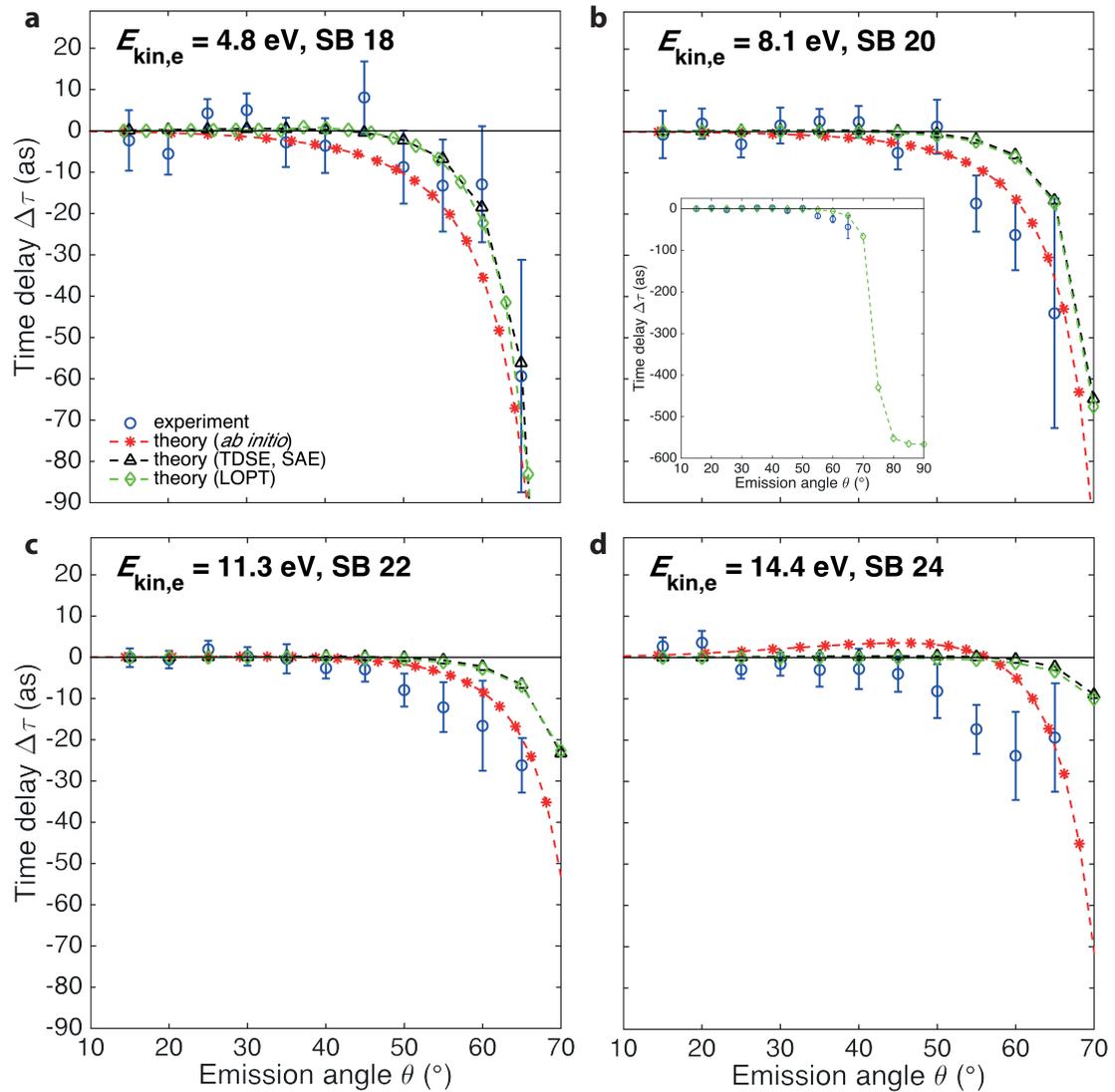

**Figure 3 | Angular dependence of photoemission time delays in helium for different electron kinetic energies. a-d,** For all electron kinetic energies, referenced by the sidebands (SBs) of the harmonic spectrum of the attosecond XUV pulse train, the experimentally retrieved atomic delay (blue data points with error bars) is shown as a function of the emission angle $\theta$, following the procedure described in Fig. 2. For example, a delay at 15° is understood as the delay between electrons emitted at angles between 0° and 30° (reference) and electrons emitted at angles between 10° and 15°. As a comparison the corresponding theoretical predictions are also included in the graphs comprising an *ab initio* simulation (red dashed line with asterisks), a calculation solving the time-dependent Schrödinger equation (TDSE) within the single active electron (SAE) approximation (black dashed line with triangles) and lowest-order perturbation-theory (LOPT) (green dashed line with diamonds). The different theories are in very good agreement and reproduce the experimental data well. The inset in (**b**) shows the typical

behavior of the angle-dependent delay predicted by LOPT for an angular range up to 90°. As a consequence of the node in the *d*-wave, at large emission angles $\theta$ the delay changes significantly.

While the SB beating at small angles is clearly visible even in the energy resolved spectrum (Fig. 2a), it is barely discernible at large angles (Fig. 2c). When the SB signals are integrated in energy, however, the characteristic oscillations with periodicity $2\omega_{IR}$ appear for both angular ranges (Fig. 2b), and thus a clear angle-dependent delay $\Delta\tau$ can be extracted. This is the delay between electrons emitted at angles between 0° and 30° (reference) and electrons emitted into a specific hollow cone between $\theta$ and $\theta+\Delta\theta$ (Fig. 1a). The accuracy of the fit decreases at larger angles due to the smaller count rate thus resulting in larger error bars. Note that the angular range of the reference has been chosen to be as wide as 30º in order to improve its signal-to-noise ratio and thus to minimize the error in the relative phase retrieval.

The measured angle-resolved photoemission time delays relative to the zero emission angles are shown with error bars in Fig. 3 for four consecutive sidebands, SB 18-SB 24. For all sidebands, the measurements deviate significantly from zero for angles larger than 50°. The largest anisotropy is recorded for the lowest sideband, but it is statistically significant in all cases.

To validate and explain the experimental observations, we used different theoretical models. First, we performed *ab initio* simulations solving the full dimensional time-dependent Schrödinger equation (TDSE) by using a nearly exact method[30], which takes into account both of the electrons in He. The method is based on a B-spline close-coupling representation of the ionization channels, with a full configuration interaction pseudochannel to account for short-range correlation[31]. Starting from the ground state, the atomic wave function evolves under the action of a second-order exponential time-step operator that accounts for the interaction with the external pulses[32]. At the end of the pulse, the resulting wave packet is projected on the scattering states of the atom, thus obtaining the differential asymptotic distribution of the photoelectrons[30,32]. This method reproduces accurately the atomic dynamics and the pulses used for the simulation can be tailored to reproduce those employed in the actual experiments. To all practical purposes, therefore, the *ab initio* results are expected to be a faithful numerical replica of the real experiment. Figure 4 shows a comparison between the time-delay integrated photoelectron spectra measured in the experiment for a moderately weak ($3 \times 10^{11}$ W/cm$^2$) IR probe pulse with a center wavelength of 780 nm and the spectrum computed *ab initio* using pulse parameters that match the experimental ones. Figure 3 shows the comparison of the time delays $\Delta \tau$ for the energy integrated SB signals. The results of the *ab initio* calculations are in quantitative agreement with the measurement. For SB 18 the experimental data slightly deviate from the theoretical estimates as compared to the other SBs. We attribute these deviations to the low intensity of SB 18 and consequently to a noisier signal, reflected also in larger error bars. Nevertheless, for the considered angular range, the discrepancy between experimental data and the theory curves is not larger than 10-15 as, which we consider fairly acceptable given the complexity of experiment and theory.

As it is known, the finite duration of the pulses gives rise to harmonics with a finite width, whose tail partly overlaps with the sidebands in the energy-resolved photoelectron spectrum. This effect, which is entirely negligible for angle-integrated measurements, is noticeable in angle-resolved measurements. To make sure that the observed anisotropy is still present in absence of any spectral overlap from the harmonics, we repeated the *ab initio* simulations with long laser pulses and compared the result with the prediction of an independent lowest-order perturbation-theory (LOPT) calculation that assumes infinitely long pulses.

Angle-resolved atomic delay LOPT calculations are performed using correlated two-photon (XUV+IR) matrix elements on an exterior complex scaled basis set[20]. This approach accounts for correlation effects in the single-XUV-photon absorption with the screening of the remaining electron in the residual ion. The interaction with the IR field, however, is treated as an uncorrelated transition. Hence, if correlation effects in the final state become significant, LOPT may fail in explaining the observed experimental results. The two calculations are in excellent agreement and the prediction of the latter is shown in Fig. 3 (represented as a dashed green line with diamonds). Even if the time-delay anisotropy of this second set of calculations is smaller than before, the effect is still clearly visible, and in particular, the sharp drop around 50° is reproduced.

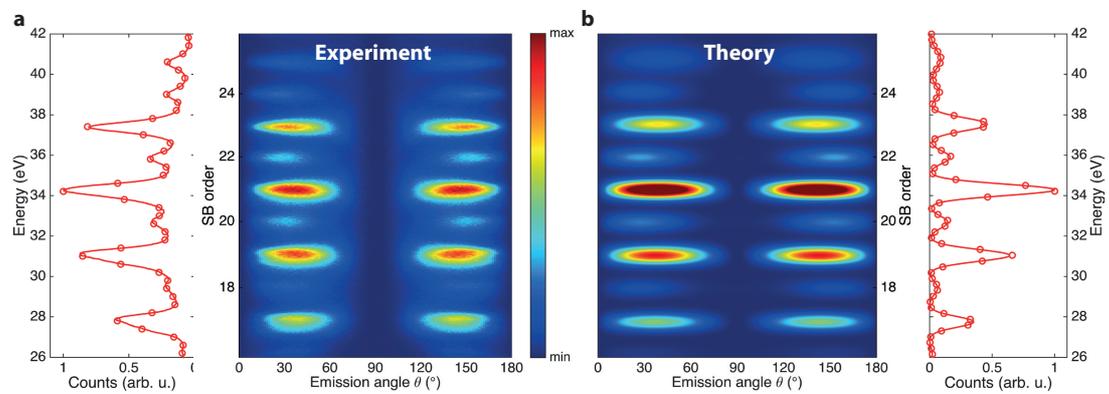

**Figure 4 | Comparison between experiment and *ab initio* simulation. a,** Experimental data. **b,** Results of the *ab initio* calculation. The 2D plots show the delay-integrated photoelectron spectrum as a function of the emission angle *θ*, defined in Fig. 1a. On the left and right hand side of (**a**) and (**b**), respectively, the angle-integrated projections for experiment and theory highlight the presence of four SBs comprising SB 18 to SB 24.

Within LOPT, the anisotropy of the time delay can be explained with an analytical description, which gives better physical insight into this anisotropy. As described in the introduction, if two linearly polarized photons are involved in the ionization of He, two different final states become accessible, represented by an *s*- or a *d*-wave. The angular shape of each final state can be described by a distinct spherical harmonic, $Y_l^m$, with *l* (*m*) representing the orbital angular momentum (magnetic) quantum number. While the $Y_0^0$ spherical harmonic representing the *s*-wave is isotropic, the $Y_2^0$ spherical harmonic associated to the *d*-wave exhibits a node at the magic angle of 54.7°. Therefore, the interference between the transitions in the continuum mediated by the IR pulse is expected to lead to an angular dependence of the atomic time delay. The variation of the delay is expected to become particularly pronounced when the emission direction of the photoelectrons with respect to the XUV and IR polarization axis approaches 60°. We can parametrize the observed angle-dependent delay in the special case of He as follows:

$$\Delta \tau = 1/2\omega_{IR} \cdot \arg\left((1+T^-)/(1+T^+)\right), \quad (1)$$

with $T^\pm = \sqrt{4\pi} \cdot c_\pm^{ds} e^{i\phi_\pm^{ds}} Y_2^0(\theta,0)$. Here, $c_\pm^{ds} = |A_\pm^d / A_\pm^s|$ and $\phi_\pm^{ds} = \arg(A_\pm^d / A_\pm^s)$ are the absolute values and phases of the two-photon transition amplitudes representing the four quantum paths *s* → *p* → *s* (+/-) and *s* → *p* → *d* (+/-). The symbol (+) indicates the transition involving the absorption of an IR photo and (-) represents the transition, which involves the emission of an IR photon.

The inset in Fig. 3b shows the behavior of the angle-dependent delay predicted by LOPT up to 90°. As soon as the magic angle of approximately *θ*=54.7° is reached, the *d*-wave changes sign and therefore exhibits a significant change in delay, which can be as large as 600 attoseconds, outside the experimentally accessible angular range.

The lack of experimental data for large angles in case of the He target prevents us from an accurate determination of the parameters $c_\pm^{ds}$, $\phi_\pm^{ds}$. Other targets may have a smaller critical angle, which then would be more easily accessible with our experimental setup. In that case a robust parametrization of the time-delay angular dependence could be obtained thus providing a simple analytical way to estimate the degree of anisotropy in such kind of measurements.

In contrast to the other SBs, theory predicts a slight positive delay for SB 24 at angles smaller than about 55°, a trend that is not observed in the experimental data (Fig. 3). We attribute this effect to the spectral overlap of SB 24 with its two neighboring harmonics 23 and 25 for which the difference in intensity is the largest (Fig. 1c).

What is the physical origin of this angular modulation in the relative phase between the *s*- and the *d* waves? Is it due to correlation, is it essentially a multi-electron effect, or is it present also for hydrogenic systems? To answer these questions, we conducted two additional test calculations. In one, we solved the TDSE using a SAE model, following a strategy tested in previous studies[33]. This model is known to reproduce well both the ionization potential of He and the one-photon ionization cross section of the atom. However, by construction, it does not account for any correlation effects between the two electrons. The predictions of this model are shown in Fig. 3 as triangles following a black dashed line and are matching the values calculated with LOPT. We conclude, therefore, that correlation has only a minor influence on the observed anisotropy.

After revisiting the phase-shifts induced by the IR field in atomic hydrogen[24], we found evidence for small phase differences at low electron kinetic energies that depend on the final angular momentum of the electron. Surprisingly, these phases mostly cancel in angle-integrated experiments and have so far not been studied in greater detail. This opens up the question if the anisotropy, similar to that found in He, can be expected in atomic hydrogen where multi-electron effects are absent. To investigate this, we made additional simulations with atomic hydrogen. Figure 5 shows a comparison of the results obtained from *ab initio* calculations for the angle-dependent time delay of He (red solid lines) and atomic hydrogen (blue and black solid lines) for a sideband centered in both cases at a photoelectron kinetic energy $E_{\text{kin,e}}$ of 4 eV, which is close to the energy of SB 18 in Fig. 3.

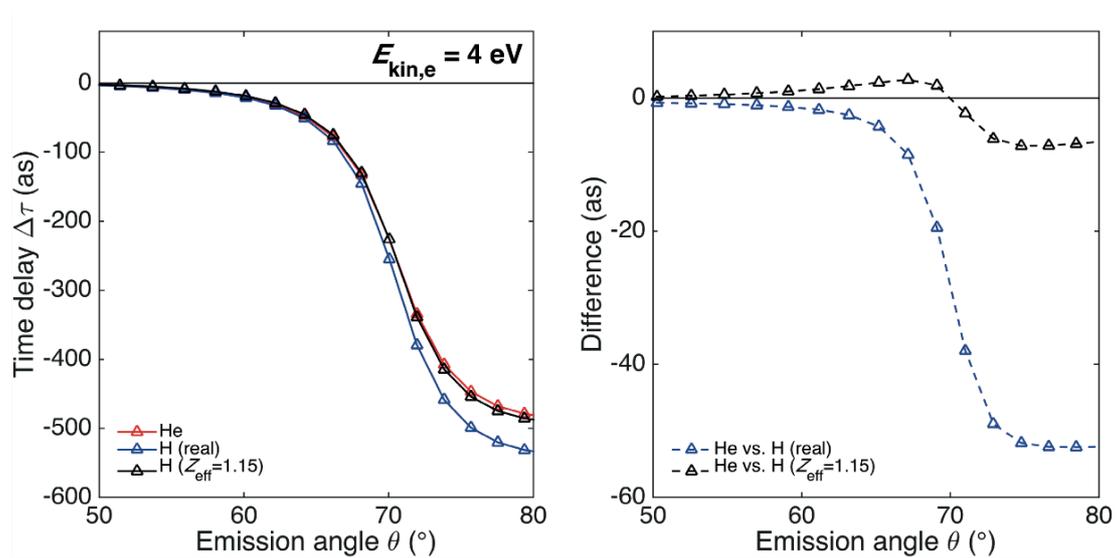

**Figure 5 | Comparison between helium and atomic hydrogen.** Comparison between the results obtained from *ab initio* calculations for the angle-dependent time delay (left panel) and their differences (right panel) of He (red solid lines), atomic hydrogen (blue solid lines) and artificial atomic hydrogen with an effective charge of $Z_{\text{eff}} = 1.15$ (black solid line). The last case should mimic a He atom in a simple single active electron (SAE) approximation in a screened potential. In order to reduce the effect of spectral overlap induced by the neighboring harmonics, the energy integration window has been chosen to be narrower than for the experimental data thus resulting in slightly smaller delays.

The trend of decreasing time delays with increasing emission angle is similar but not exactly the same for both He and atomic hydrogen (Fig. 5). The reason for this difference, which can be as large as 60 attoseconds, is due to the remaining electron in the He$^+$ (1s) parent ion, which modifies the effective potential seen by the departing photoelectron. Indeed, as can be seen in Fig. 5, this effect can be largely accounted for by using a fictitious hydrogen atom with an effective charge $Z_{eff} = 1.15$, which mimics the screening of the He nuclear charge by the 1*s* electron.

## Conclusions and Outlook

We have provided the first experimental evidence of an angular dependence in the measurement of photoemission time delays. These measurements are based on single photoionization, realized in a non-sequential pump-probe scheme where the extreme ultraviolet (XUV) attosecond pump pulse ionizes the target system and an infrared (IR) probe pulse interacts with the liberated electrons. We have observed an angular dependence even when the single-photon emission delay is rigorously isotropic. This photoemission angular dependence results from the interference between two different final quantum states accessible in two-photon processes. With the help of state-of-the-art theories, we show that the observed time delay anisotropy in He is due to the effect of the spherical ionic potential on the outgoing photoelectron during the probe stage. In particular, at variance with atomic hydrogen, where anisotropies are also expected, the remaining electron in the He$^+$ (1s) parent ion has a noticeable effect on the observed anisotropy, thus pointing out the potential of this technique to investigate multi-electron effects from angularly resolved time delays.

The above conclusions apply to most attosecond measurement techniques, such as streaking and RABBITT. This knowledge may shed new light on previous experiments performed in gaseous[3,6] and condensed matter systems[5], where the angular dependence of the measured time delays was not always taken into account and where in most cases SAE approximations have been used.

## Acknowledgement

S.H, C.C, L.G. and U.K. acknowledge support by the ERC advanced grant ERC-2012-ADG_20120216 within the seventh framework program of the European Union and by the NCCR MUST, funded by the Swiss National Science Foundation. M.L. acknowledges support from the ETH Zurich Postdoctoral Fellowship Program.

Á.J.G., L.A. and F.M. acknowledge the support from the European Research Council under the ERC grant no. 290853 XCHEM, from the European COST Action CM1204 XLIC, the MINECO Project FIS2013-42002-R, the ERA-Chemistry Project PIM2010EEC- 00751, and the European Grant MC-ITN CORINF. Calculations were performed at the Centro de Computación Científica of the Universidad Autónoma de Madrid (CC-UAM) and the Barcelona Supercomputing Center (BSC).

I.I. and A.S.K. acknowledge support of the Australian Research Council (DP120101805) and the use of the National Computational Infrastructure Facility.

J.M.D. and E.L. acknowledge support from the Swedish Research Council (VR). Moreover, this research was supported in part by the National Science Foundation under Grant No. NSF PHY11-25915.